\documentclass[a4paper,fleqn,usenatbib]{mnras}

\usepackage{newtxtext,newtxmath}

\usepackage[T1]{fontenc}
\usepackage{ae,aecompl}

\usepackage{graphicx}	
\usepackage{amsmath}	
\usepackage{amssymb}	
\usepackage{pdflscape}	
\usepackage{verbatim}
\usepackage{color}
\usepackage{pgfplotstable}
\usepackage{booktabs}
\usepackage{rotating}

\graphicspath{{/Users/ireagu/Documents/A2151_pap/fig/}}

\title[Spectroscopic analysis of Abell\,2151]{Deep spectroscopy of nearby galaxy clusters: II.
The Hercules cluster.}

\author[I. Agulli et al.]{I. Agulli,$^{1,2,3,4}$ \thanks{E-mail: ireagu@gmail.com}, J. A. L. Aguerri,$^{1,2}$ \thanks{E-mail: jalfonso@iac.es}, A. Diaferio$^{3,4}$ \newauthor
L. Dominguez Palmero,$^{5,1}$, R. S\'anchez-Janssen,$^{6}$ \\
$^1$ Instituto de Astrofisica de Canarias, C/ Via Lactea s/n, E-38205 La Laguna, Tenerife, Spain, \\
$^2$ Departamento de Astrofisica, Universidad de La Laguna, E-38206 La Laguna, Tenerife, Spain,\\
$^3$ Dipartimento di Fisica, Universit\`a di Torino, Via P. Giuria 1, I-10125 Torino, Italy\\
$^4$ Istituto Nazionale di Fisica Nucleare (INFN), sezione di Torino, Via P. Giuria 1, I-10125  Torino, Italy\\
$^5$ Isaac Newton Group of Telescopes, Apartado 321, E-38700 Santa Cruz de La Palma, Canary Islands, Spain\\
$^6$ STFC UK Astronomy Technology Centre, Royal Observatory, Blackford Hill, Edinburgh, EH9 3HJ, UK\\
}

\date{Accepted 2017 February 9. Received 2017 February 8; in original form 2016 August 5.}

\pubyear{2016}

\begin{document}
\label{firstpage}
\pagerange{\pageref{firstpage}--\pageref{lastpage}}
\maketitle

\begin{abstract}
We carried out the deep spectroscopic observations of the nearby cluster A\,2151 with AF2/WYFFOS@WHT. The caustic technique enables us to identify 360 members brighter than $M_r = -16$ and within 1.3$R_{200}$. We separated the members into subsamples according to photometrical and dynamical properties such as colour, local environment and infall time. The completeness of the catalogue and our large sample allow us to analyse the velocity dispersion and the luminosity functions of the identified populations. We found evidence of a cluster still in its collapsing phase. The LF of the red population of A\,2151 shows a deficit of dwarf red galaxies. 
Moreover, the normalized LFs of the red and blue populations of A\,2151 are comparable to the red and blue LFs of the field, even if the blue galaxies start dominating one magnitude fainter and the red LF is well represented by a single Schechter function rather than a double Schechter function. We discuss how the evolution of cluster galaxies depends on their mass: bright and intermediate galaxies are mainly affected by dynamical friction and internal/mass quenching, while the evolution of dwarfs is driven by environmental processes which need time and a hostile cluster environment to remove the gas reservoirs and halt the star formation.   
\end{abstract}




\section{Introduction}
In the current paradigm of galaxy formation, baryons are embedded in dark matter haloes which grow through merger events across cosmic time. The properties and subsequent evolution of baryons depend on the halo they live in. The largest halos observed in the Universe are the galaxy clusters which forms via accretion and mergers of smaller objects.   

Since the seventies there are studies showing that the galaxy properties depend on the environment, with a low fraction of star forming galaxies in high density environments \citep[e.g.][]{oemler1974,dressler1980,fasano2000,fasano2015,lewis2002}. Indeed, galaxies embedded in the cluster environment suffer of many physical processes affecting their stellar and gas component, such as ram pressure stripping, harassment and tidal interactions \citep[e.g.][]{wr1978,gunn1972,moore1996,quilis2000,bekki2002,aguerri2009,kawata2008}. However, disentangling the time scales and the processes affecting the star formation history of galaxies is still an active debate.

The faint galaxies are usually defined as dwarfs and have $r-$band absolute magnitudes $\geq -18$ ($M_r \gtrsim M_r^* + 4$). This population is the most abundant type of galaxies in the Universe and is highly affected by external processes because of their shallow potential \citep[][]{lisker2012}. In particular, their star formation histories are strongly affected by the environment \citep[e.g.][]{peng2010,peng2015,davies2016}. A fraction of them could also be the product of strong tidal transformation in clusters \citep[e.g.][]{aguerri2016}. However, deep spectroscopic studies of this population in different environments are only a recent achievement and are usually limited to the nearby Universe. Therefore, even if extremely interesting thanks to the constrains on their evolution, the spectroscopic study of dwarfs in young and/or forming clusters remain challenging.  

One of the largest and most massive structures in the Local Universe is Hercules Supercluster, composed by Abell 2151 (the Hercules cluster and hereafter A\,2151), Abell 2147 and Abell 2152 \citep{chincarini1981,barmby1998}. It is probably bound as a whole object and, in particular, the kinematical analysis of \cite{barmby1998} reveals that A\,2151 and Abell 2147 are bound to each other. A\,2151 is a nearby (z = 0.0367), irregular and spiral-rich cluster \citep[$\sim 50 \%$][]{giovanelli1985}. The lack of hydrogen deficiency in the spiral population \citep{giovanelli1985,dickey1997}, together with the bumpy distribution of the hot intracluster gas and its low X-ray flux \citep{magri1988,huang1996} and the presence of at least three distinct substructures \citep{bird1995}, are strong evidence of a cluster still in its merging epoch. Moreover, \cite{cedres2009} studied the H$\alpha$ distribution in A\,2151 and suggested a dependence of star formation on both the global environment and the merger history. Therefore, this cluster shows different properties compared to very massive and formed clusters such as Coma and A\,1367. Indeed, these are all evidence of a relatively unevolved and young cluster. 

Therefore, the study of A\,2151 with deep spectroscopic observations can provide interesting constrains on the evolution of the dwarf population. The dynamical properties and the luminosity function (LF) are powerful observables for this kind of studies, and different subsets depending on the stellar colour, the luminosity and/or the dynamical properties of A\,2151 members can be analysed. Moreover, their comparison with literature studies of different environments lead to the understanding of the physical mechanisms involved. 

Photometric LFs of stacked clusters predict the presence of an upturn around $M_r \sim -18$ and a very steep faint end \cite[e.g.][]{popesso2006,lan2016}, while recent studies of photometric LFs using different constrains on the membership observed similar results with less steep faint ends \citep[e.g.][]{moretti2015,ferrarese2016}. Even if spectroscopic observations are very important to obtain accurate membership, only a handful of nearby clusters are analysed in the literature with deep spectroscopic data sets. The LFs of Virgo and Abell 2199 by \cite{rines2008} present flat faint ends with slope $\sim -1.1$ and without upturns. The Coma cluster also presents a flat LF when analysed with deep spectroscopic data \citep[][]{mobasher2003}. 

Therefore, we started a survey to obtain deep spectroscopy for nearby clusters reaching dwarf luminosities within the virial radius. The aim of this project is the analysis of the LF and the galaxy orbital properties in these clusters and we started a series of papers. In particular, we use different galaxy subsets depending on their stellar colour, their position in the cluster, their luminosity and/or their dynamical properties. 

The rich cluster Abell 85 (A\,85) was the starting point of our project. Its LF presents an upturn and the best model is a double Schechter function \citep[][]{schechter1976}. However, the faint-end slope is less steep than the photometric estimates \citep[][]{agulli2014}. Moreover, its LFs of the red and blue populations compared with the same populations of the field by \cite{blanton2005} show that the environment mainly affects the gas component halting the star formation \citep{agulli2014,agulli2016a}. Indeed, the slopes of the faint ends of the cluster and the field LFs are compatible within the observational uncertainties. Therefore, the main physical processes involved do not change significantly the masses of the galaxies. The analysis of the LFs in radial bins highlighted that the relative contribution of the two populations is similar in the cluster outskirts and that red galaxies clearly dominate the inner cluster region. Therefore, we suggested that the physical processes affecting the star formation are stronger in the cluster centre, where the density is higher \citep[][]{agulli2016a}. The analysis of the radial anisotropy profile for this cluster revealed that the blue population is formed by galaxies in less radial orbits than the remaining galaxy population and in virial equilibrium with the cluster potential (Aguerri et al., in preparation). Therefore, the blue galaxies of A\,85 and not located in substructures are not recent arrival. As a consequence, the orbital properties of the galaxies also affect their gas content and their star formation histories. 

The aim of this paper is the analysis of the different populations of A\,2151 studying their LFs and their properties. Indeed, the peculiar dynamical state of the cluster and its small distance allow a detailed analysis of the dwarf population in a merging cluster. 

We present the data set in Sec. \ref{data}. The different subsets and their properties are analysed in Sec. \ref{gc}. Section \ref{lf_sec} describes the spectroscopic LF and its study and Sec. \ref{disc} discusses the results. The summary of this analysis and the conclusions are given in Sec. \ref{concl}. 

Throughout this work we have used the cosmological parameters $H_0 = 75 \; \mathrm{km} \, \mathrm{s}^{-1} \mathrm{Mpc}^{-1}$, $\Omega _m = 0.3$ and $\Omega _{\Lambda} = 0.7$. 

\section{The data of A2151}\label{data}
\subsection{Target selection and deep AF2/WYFFOS spectroscopy}\label{tg}
\begin{figure}
\centering
  \includegraphics[width=1\linewidth]{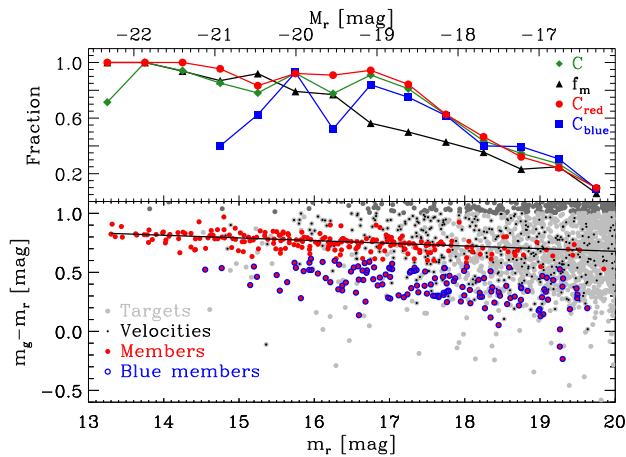}
  \caption{Lower panel: colour-magnitude diagram of the galaxies in the direction of A\,2151. Dark grey dots are the galaxies excluded from the target sample with the colour-cut. Light grey dots are the target galaxies and black points are the velocities measured. Red and blue symbols show red and blue cluster members, respectively. The solid line represents the red sequence of the cluster. Upper panel: spectroscopic completeness  ($C$, green diamonds), red ($C_\text{red}$, red dots) and blue ($C_\text{blue}$, blue square) spectroscopic completeness, and cluster member fraction ($f_\text{m}$, black triangles) as a function of $r$-band magnitude.}
\label{A2151_cmd}
\end{figure}

To carry out a deep spectroscopic coverage of the A\,2151 field, as targets, we selected the galaxies with no measured spectroscopic redshift, brighter than $m_r = 20.5$, bluer than $m_g -m_r \leq 1.0$ and within 45 arcmin from the cluster mass centre $\alpha$ (J2000): $16^\text{h} \, 05^\text{m} \, 26^\text{s}$, $\delta$ (J2000): $17^{\circ} \, 44' \, 55"$ \citep{sanchez2005}. The parent photometric catalogue we used is the Sloan Digital Sky Survey, Data Release 9 \citep[SDSS-DR9,][]{ahn2012}. Through this work, we used the modelled magnitude of SDSS-DR9 corrected for extinction. The resulting 5840 targets are shown in Fig. \ref{A2151_cmd}. The magnitude limit considered in this work is given by the properties of the spectrograph used for the observations and the colour cut is applied to minimise the background sources while matching the colour distribution of the galaxies in the nearby Universe \citep[see e.g.,][]{hogg2004,rines2008}.

We performed the observations with the fibre spectrograph AutoFiber 2/WYFFOS at the \textit{William Herschel Telescope} (AF2@WHT) during three nights (proposal C-140, 2014 May). In addition, we obtained 8 hours of service time in the same semester to complete the program (Service program SW2014a27, 2014 first semester). We collected three exposure of 1800\,s per pointing to reach a signal to noise (S/N) higher than 5 for faint galaxies. Indeed, we used the grism R158B that has a resolution of $R\,\sim\,280$ at the central wavelength $\lambda \sim 4500\, \AA$. We designed the 8 pointings to maximise the number of targets within the radius of 20 arcim, where the instrument has an optimal response. As a result we placed $\sim 90$ galaxies per pointing within that radius and a few outside. 

The data reduction of the 738 spectra observed was performed with the instrument pipeline, version 2.25 \citep{dominguez2014}. This \textsc{idl} code allows the standard reduction and the correction for the attenuation of each fibre, previously evaluated with the dome flat.

\subsection{Velocity catalogue}
\begin{figure}
\centering
 \includegraphics[width=1\linewidth]{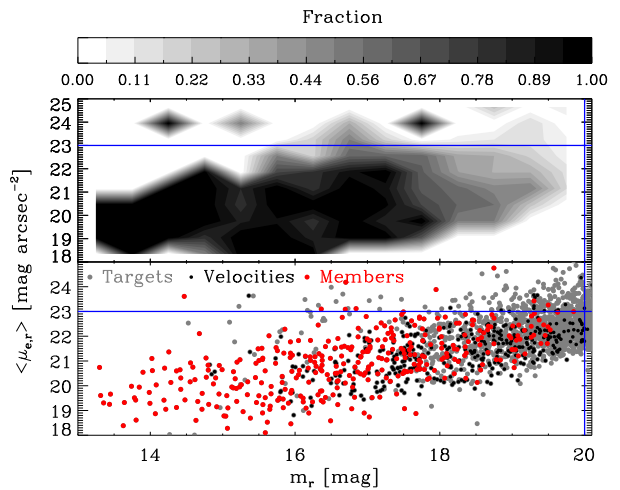}
 \caption{The upper panel shows the completeness as a function of both the apparent magnitude and the mean surface brightness using grey-scale colour levels. In the lower panel the mean surface brightness as a function of apparent magnitude is plotted: the targets with grey dots, the determined velocities with black dots and the members with red dots. The vertical and horizontal blue lines are the catalogue thresholds for the completeness.}
\label{A2151_binned_compl}
\end{figure}

To determine the recessional velocities from the observed spectra, we used the \textit{rvsao.xcsao} \textsc{iraf} task \citep{kurtz1992}. More details of this procedure can be found in \cite{kurtz1992} and \cite{tonry1979}: it basically cross-correlates a template spectrum library \citep[in this work,][]{kennicutt1992} with the observed galaxy spectrum. With this task, we determined 435 velocities, while the remaining spectra had low S/N to rely on the cross-correlation results. Even if the number of lost spectra is quite high, we could use them to put an upper limit to the mean surface brightness, $\langle \mu_{e,r} \rangle$, required for this kind of studies with AF2@WHT. Indeed, the lower panel of Fig \ref{A2151_binned_compl} clearly shows that for targets with $\langle \mu_{e,r} \rangle > 23$ mag\,arcsec$^{-2}$  almost no velocities could be identified with this setup and observing time. 

Our observational strategy was designed to repeat observations of few objects in different pointings in order to estimate the real uncertainties on the measured recessional velocities. The formal uncertainties provided by the task are smaller than the intrinsic ones \citep{bardelli1994}. Indeed, the task does not count for other sources of uncertainties such as the intrinsic resolution of the spectra and the wavelength calibration. The root mean square of the differences in the velocity measurements of the 57 repeated galaxies is 175 km\,s$^{-1}$. Moreover, we had between two and four objects per pointing with spectroscopic information in SDSS-DR9 to control the reduction process. We obtained differences in the velocities always smaller than 100 km\,s$^{-1}$.

Our new data together with spectroscopic redshifts from the literature, in particular from SDSS-DR9 and the NASA/IPAC Extragalactic Database (NED), result in 799 recessional velocities in the A\,2151 field of view, covering a radius of 0.94 Mpc from the cluster centre, of galaxies brighter than $m_r = 20.5$ and bluer than $m_g -m_r = 1.0$. We would like to point out that 362 of them are new velocities in the magnitude range of dwarf galaxies\footnote{In this study we consider dwarf the galaxies with $M_r \geq -18$ and bright those galaxies with $M_r \leq -20$.} \citep[presented the first time in][]{agulli2016b}. Fig \ref{A2151_cmd} shows the redshifts in the colour-magnitude diagram, lower panel, and the completeness of the spectroscopic dataset, $C$, in the upper panel. $C$ is defined as the ratio between the number of measured velocities, $N_z$, and the number of targets, $N_{\text{phot}}$, per each magnitude bin: $C = N_{z} / N_{\text{phot}}$, it is higher than $\sim 80\%$ for $M_r \lesssim -18.5$ and decreasing to $\sim 30 \%$ for $M_r \sim -17$. Moreover, Fig. \ref{A2151_binned_compl} shows the mean surface brightness limit of the present study: the completeness is shown as a function of both $\langle \mu_{e,r} \rangle$ and apparent magnitude. For the evaluation of observing biases we can say that the target population is well represented by the galaxies with spectroscopic information down to $m_r = 20$ mag and $\langle \mu_{e,r} \rangle \leq 23$ mag arcsec$^{-2}$.

\subsection{Cluster membership}\label{memb}
The caustic method \citep{diaferio1997,diaferio1999,serra2011} is a technique to estimate the escape velocity and the mass profile of a cluster both in its virial and infall regions. Indeed, cluster galaxies defined a region on the line-of-sight velocity--projected clustercentric distance plane that has a typical trumpet shape: decreasing amplitude, $A$, with increasing clustercentric distance, $r$. $A(r)$ can be parametrized as a function of the escape velocity and modulated by a function depending on the galaxy velocity anisotropy parameter \citep[see][]{diaferio1997}. A by-product is the cluster member identification. This method does not require the assumption of dynamical equilibrium, so it is a powerful tool to find the membership of A\,2151, due to its peculiar dynamical state. An interesting detail is that the galaxies are connected estimating their binding energy and constructing a binary tree. Then, the main group of the binary tree is used to calculate the centre, the velocity dispersion and the radius of the cluster. With these data, the redshift diagram is created, the galaxies are weighted with a kernel function and the caustic curves are defined. The mass profile of the cluster is estimated from the caustics. Note that the caustics are symmetric by definition. As \cite{serra2013} showed, the membership obtained from the caustic is more accurate at larger radii than the one from only the binary tree. However, both have a very small interloper contamination within the virial radius, 2\% for the former and 3\% for the latter. 

\begin{figure}
\centering
  \includegraphics[width=1\linewidth]{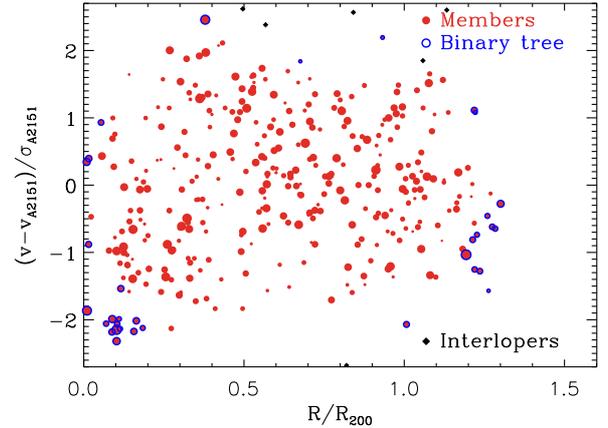}
  \caption{Line-of-sight velocity -- projected clustercentric distance plane for the A\,2151 members (red dots) and the interlopers (black diamonds). The dimensions of the red dots scale with the magnitude: bright (faint) -- big (small) dots. Blue circles identified the members determined only by the binary tree and included in the analysis.}
\label{A2151_caust}
\end{figure}

\begin{figure}
\centering
 \includegraphics[width=1\linewidth]{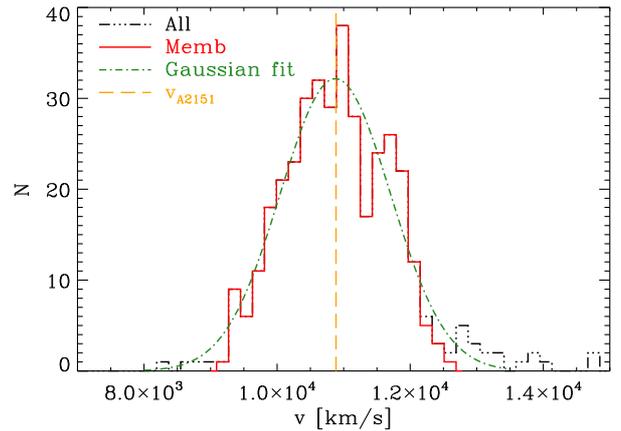}
 \caption{The velocity histogram of the galaxies within $\sim 5 \sigma_\text{c}$ is presented in black dash-dot-dot-dotted line, while the member one is presented with a solid red line. The green dash-dotted line is the Gaussian fit to the velocity histogram of the members. The vertical dashed orange line shows the mean cluster velocity $v_\text{c}$.}
\label{A2151_v_hist}
\end{figure}

In the case of A\,2151 we decided to use the membership from the binary tree obtaining 360 cluster members\footnote{The data will be published on  VizieR.} (see Fig.~\ref{A2151_caust}). 
Figure \ref{A2151_v_hist} presents the histogram of the velocities for the members together with the background and foreground velocities within $5\sigma_\text{c}$. 

\begin{table}
\begin{center}
\begin{tabular}{cc}
\hline
Members		&360\\
$\alpha_\text{c}$ (J2000)  &$16^{\text{h}} \, 05^{\text{m}} \, 23^{\text{s}}.369$\\
$\delta_\text{c}$ (J2000)	&$17^{\circ} \, 45' \, 04".33$\\
$v_{\text{c}}$ 			& 10885 km\,s$^{-1}$\\
$\sigma_{\text{c}}$ 		& 704  \, [km\,s$^{-1}$]\\
$R_{200}$			&$1.45 \,$Mpc\\
$M_{200}$			& $4.00 \pm 0.4 \times 10^{14} \; $M$_{\odot}$\\
\hline
\end{tabular}
\caption{Properties of A\,2151 obtained from the caustic method. The subsctript $_\text{c}$ refers to the cluster. \label{A2151_prop}}
\end{center}
\end{table}

The cluster properties obtained with the caustic technique are summarised in Tab. \ref{A2151_prop}. In particular, we present the cluster centre, the cluster recessional velocity and velocity dispersion ($v_\text{c}, \; \sigma_\text{c}$), and the virial mass and radius ($M_{200}, \; R_{200}$). According to these results, we cover $\sim1.3 \,R_{200}$ with 360 members.

A\,2151 does not have a BCG formed yet and our catalogue is different with respect to previous studies. As the caustic method can also find the centre of the galaxy distribution, we evaluated the centre and we will use it for further analysis. This cluster centre lies 80 kpc away from the cluster mass centre estimated by \cite{sanchez2005}. Moreover, the values of $v_\text{c}$ and $\sigma_\text{c}$ obtained with our catalogue are in general agreement with literature results for A\,2151 \citep{barmby1998,girardi1998,struble1999}.

Fig \ref{A2151_cmd} shows the member fraction, $f_{\text{m}}$, of A\,2151 in the upper panel. It is defined as $f_{\text{m}} = N_{\text{m}} / N_z$, being $N_{\text{m}}$ the number of members. We notice that $f_{\text{m}}$ strongly depends on the luminosity and that we reach $m_r \sim 19.5$.

\section{Galaxy classification}\label{gc}
We classified the members into different populations in order to study their physical properties. We will focus on the dynamics and the LF. The details are give in the following sections.

\subsection{Red and blue members}\label{red_blue}
We defined red and blue members as described in \cite{agulli2016b}. They presented a detailed analysis of these two populations. Indeed, A\,2151 is a peculiar nearby cluster, with a majority of blue dwarf galaxies: only 36 per cent of the galaxies with $M_r \leq -18$ are located on the red sequence. 

The best linear model to fit the red sequence is $(m_g -m_r)_\text{RS} = -0.02 m_r + 1.13$ with $\sigma = 0.053$ and was obtained with all the members brighter than $m_r \sim 17$ and with $(m_g -m_r) > 0.6$. The members are classified as blue if $(m_g -m_r) < (m_g -m_r)_\text{RS} - 3 \sigma_\text{RS}$ and they result to be the 39\ per cent of the sample spanning the full magnitude range. Fig \ref{A2151_cmd} shows the red-sequence fit and the two populations in the colour magnitude diagram. Note that the paucity of red dwarfs is not due to a bias in the observations. A visual inspection of Fig \ref{A2151_cmd} confirm that we measured velocities in the faint region of the red sequence, but the galaxies resulted to be background objects. Moreover, the completeness of the two populations, $C_{\text{red}}$ and $C_{\text{blue}}$, defined as the ratio between the velocity of red (blue) galaxies measured with respect to the red (blue) targets, is presented in the upper part of Fig. \ref{A2151_cmd} and the two types of galaxies have been equally covered with the observations \citep[see][for details]{agulli2016b}. 

\subsection{Virialized and non-virialized members}\label{sec_sub}
The X-ray analysis and the comparison with its optical counterparts suggest the presence of substructures in A\,2151 \citep{bird1995,huang1996}. Our catalogue includes new confirmed members and in particular 94 dwarfs. The first evidence of the presence of substructure or a merging epoch is a velocity distribution which deviates from a Gaussian distribution. Therefore, we applied the \textsc{rostat} statistical analysis \citep[see][]{beers1990} that performs different statistical test evaluating the location and the scale of the velocity distribution with classical estimators as \textit{Skewness}, \textit{S}, and \textit{Kurtosis}, \textit{K}, and with \textit{Tail Index}, \textit{TI}, and \textit{Asymmetric Index}, \textit{AI}, that are more robust against outliers, contaminations and non-Gaussinaity of the data set. The results are summarised in Tab. \ref{A2151_rostat}. 

\begin{table}
\begin{center}
\begin{tabular}{ccc}
\hline
\hline
Test		&Value		&Rejection probability\\
\hline
\textit{S}		&-0.03	&22\%\\
\textit{K}		&-0.75	&99\%\\
\textit{TI}		&1.71	&93\%\\
\textit{AI}		&-0.22	&33\%\\
\hline
\end{tabular}
\caption{Results from the \textsc{rostat} analysis on the velocity distribution of A\,2151 and their rejection probability obtained with 10,000 MC simulations.\label{A2151_rostat}}
\end{center}
\end{table}

Fig \ref{A2151_v_hist} shows the velocity histogram for A\,2151 with a simple Gaussian fit. A visual inspection suggests that this distribution is not deviating from a Gaussian distribution. Moreover, the Kolmogorov--Smirnov test gives that the distribution of the member velocities and the fitted Gaussian are not statistically different. Assuming a Gaussian distribution and evaluating the probability density distribution, $f(x)$ with 10,000  Monte Carlo (MC) simulations, we defined the rejection probability, $RP$, as:
\begin{equation}\label{rej_prob_eq}
RP \equiv 1 - \int_{-a}^{a} f(x)\text{d}x
\end{equation}
where $|a|$ is the value obtained for our velocity distribution. According to \textit{K} and \textit{TI}, the velocity distribution of A\,2151 is statistically different from a Gaussian. 
 Therefore, these tests suggest the presence of substructures in A\,2151. 

\begin{figure*}
\centering
 \includegraphics[width=1\linewidth]{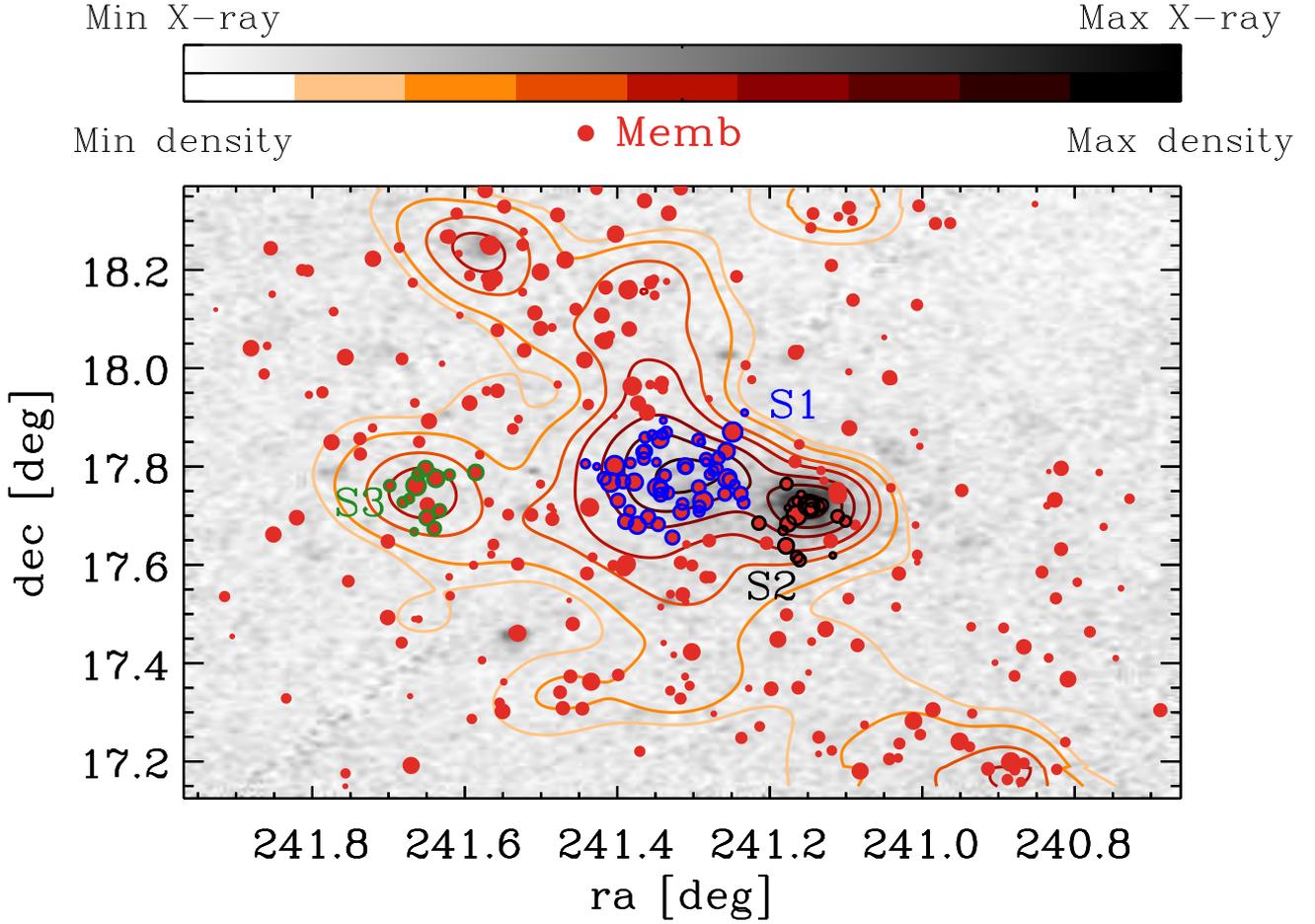}
 \caption{In this plot we show the X-ray emission map from \textit{ROSAT} (grey scale) and the heat-scale contour levels represent the density obtained with \textsc{2d-dedica} algorithm with arbitrary scales. The red dots are the members of A\,2151 with the size scaled according to the magnitude (larger size means brighter magnitude). The blue, black and green circles identify the galaxies belonging to the substructures S1, S2 and S3 respectively.}
\label{A2151_xray_cfr}
\end{figure*}

We performed a 2D-test that evaluates the local density of the members: \textsc{2d-dedica} \citep[see][]{pisani1993}. The contour levels obtained are presented in Fig \ref{A2151_xray_cfr}. Four density peaks are identified: one is located in the cluster centre, one to the east, one to the west and a last one to the north-east. The same figure shows the X-ray emission map from \textit{ROSAT} and a visual inspection evidences that the high density regions coincide with the X-ray peaks of emission. In particular, they correspond to the four peaks identified by \cite{bird1995}.

To identify the galaxies belonging to the substructures, we used the caustic method as described in \cite{yu2015}. The algorithm gave 33 per cent of the members in the substructures. Fig \ref{A2151_xray_cfr} shows the members of A\,2151 and the galaxies found in substructure by using this technique on top of the X-ray emission map and the galaxy density peaks. This figure points out that the identified substructures coincide with three of the galaxy density peaks. In particular, S1 with the central peak and S2 and S3 with the western and eastern ones, respectively.
There are no galaxies considered in substructures which correspond to the north-east density peak. However, this region is near the border, so the presence of a substructure can be biased by the edge effect. The main characteristics of S1, S2 and S3 are presented in Table \ref{A2151_sub_prop}.

\subsection{Early and recent infall members}
\cite{oman2013} proposed a line in the phase-space that estimates the infall time of the members:
\begin{equation}\label{oman_eq}
| \frac{v-v_\text{c}}{\sigma_\text{c}} |= -\frac{4}{3}\frac{R}{R_\text{vir}}+2 , 
\end{equation}
 where $v$ are the recessional velocities of the galaxies, R is the projected clustercentric distance and $R_\text{vir}$ is the cluster virial radius. We assumed $R_{200} \sim R_\text{vir}$. Using simulations, they found that a significant fraction of the galaxies above this line fell into the cluster during the last 1Gyr. When applied to our cluster, we found 31 per cent of the members in this region. For simplicity we will call these galaxies \textit{recent-infall} ($\tau_\text{inf} <1$\,Gyr) and the members below the line \textit{early-infall} ($\tau_\text{inf} > 1$\,Gyr).
 
 \begin{figure*}
\centering
 \includegraphics[width=1\linewidth]{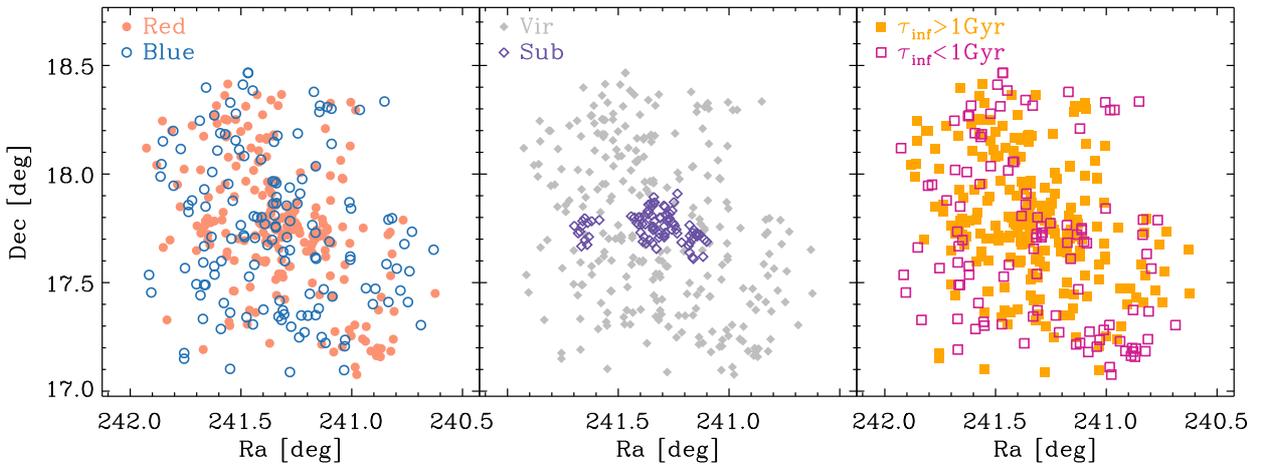}
 \caption{We present here the positions of the different populations on the sky. In the left-hand plot there are the red (red dots) and blue (blue circles) members, in the middle one the members in substructure regions (empty purple diamonds) or not (filled grey diamonds) are presented and in the right-hand one there are the members that fell in the cluster more (filled orange squares) or less (empty magenta squares) than 1 Gyr ago, according to Eq.~\ref{oman_eq}.}
\label{A2151_ra_dec_pop}
\end{figure*}

\begin{figure*}
\centering
 \includegraphics[width=1\linewidth]{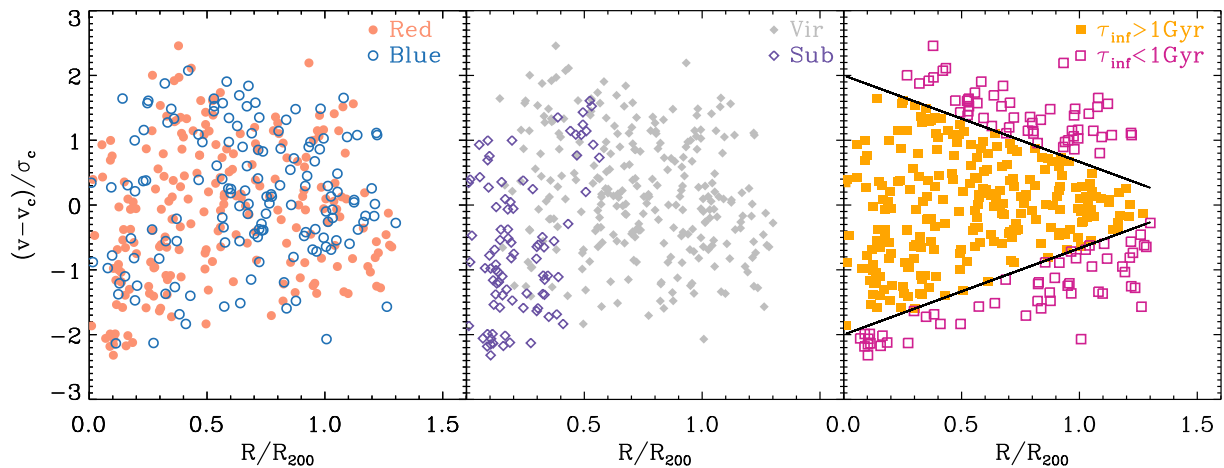}
 \caption{The phase-space of the different populations is shown for the same distributions, and with the symbol and colour codes of Fig. \ref{A2151_ra_dec_pop}. The solid black lines in the right-hand panel are obtained from Eq.~\ref{oman_eq}}
\label{A2151_caustic_pop}
\end{figure*}

\begin{figure*}
\centering
 \includegraphics[width=1\linewidth]{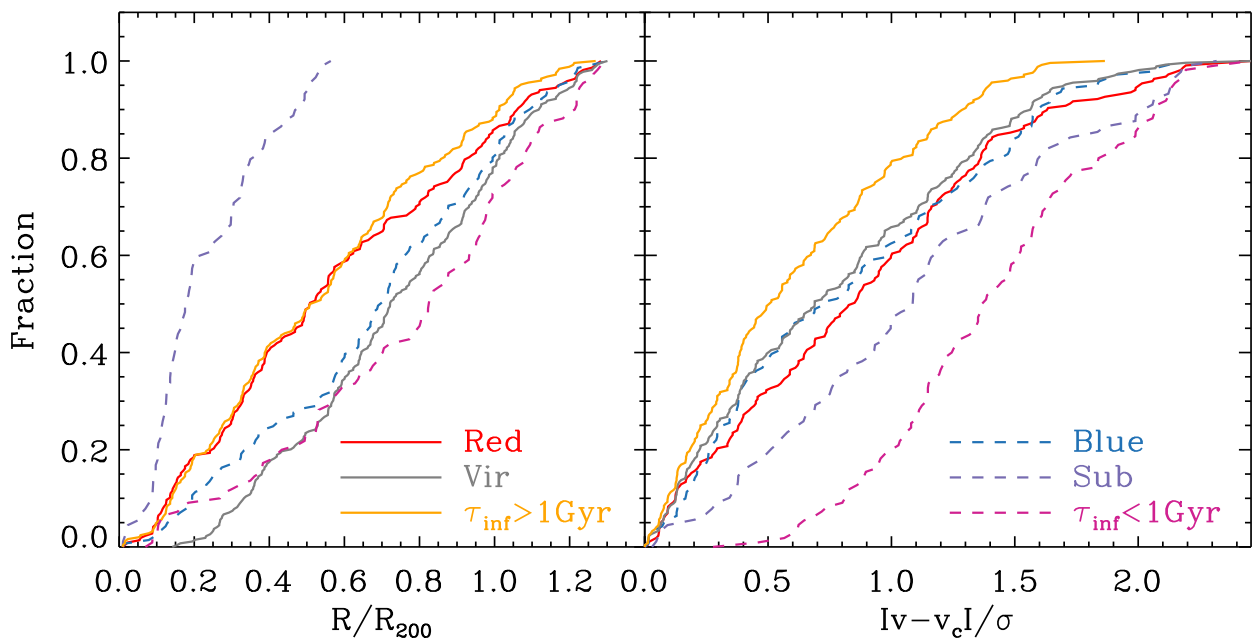}
 \caption{We present the cumulative distribution function for the different types of galaxies as a function of the radius in the left-hand panel and of the recessional velocity in the right-hand panel. The colour code is the same as in Fig. \ref{A2151_ra_dec_pop}, with the most numerous populations as solid lines and the others in dashed lines.}
\label{A2151_acum_distr_func}
\end{figure*}

\begin{table}
\begin{center}
\begin{tabular}{ccccc}
\hline
\hline
Mag range	&$N_\text{m,sub}$ 	&$\langle R/R_{200} \rangle$	&$\langle v \rangle$ [km/s]	&$\sigma$ [km/s]\\
\hline
S1		&57		&0.13	&10299		&711\\
S2		&20		&0.33	&10116		&441\\
S3		&13		&0.49	&11657		&258\\
\hline
\end{tabular}
\caption{Dynamical properties of the three substructures identified in A\,2151. The size are given in units of $R_{200}$ of A\,2151. \label{A2151_sub_prop}}
\end{center}
\end{table}

\subsection{Dynamical analysis of the galaxy populations}
Figures~\ref{A2151_ra_dec_pop} and \ref{A2151_caustic_pop} present  the spatial and phase-space distribution of the populations described above. Figure~\ref{A2151_acum_distr_func} shows the cumulative distribution function of $R/R_{200}$ and $|v-v_\text{c}|/\sigma_\text{c}$ for the same types of galaxies. Moreover, their mean clustercentric distance, mean velocity and velocity dispersion are listed in Table~\ref{A2151_pop_prop} together with the same values for the bright and dwarf galaxies of each population.

\begin{table}
\begin{center}
\begin{tabular}{ccccc}
\hline
\hline
Mag range	&$N_\text{m,pop}$ 	&$\langle R/R_{200} \rangle$	&$\langle v \rangle$ [km/s]	&$\sigma$ [km/s]\\
\hline
\multicolumn{5}{c}{Red}\\[2pt]
$-23 < M_r \leq -16$		&218		&0.57	&10816		&764\\
$-23 < M_r \leq -20$		&86		&0.59	&10927		&787\\
$-18 \leq M_r \leq -16$	&31		&0.52	&10735		&663\\[3pt]
\hline
\multicolumn{5}{c}{Blue}\\[2pt]
$-23 < M_r \leq -16$		&142		&0.66	&10965		&730\\
$-23 < M_r \leq -20$		&10		&0.69	&11094		&411\\
$-18 \leq M_r \leq -16$	&59		&0.63	&11049		&774\\[3pt]
\hline
\multicolumn{5}{c}{Virialized}\\[2pt]
$-23 < M_r \leq -16$		&270		&0.74	&11014		&683\\
$-23 < M_r \leq -20$		&72		&0.72	&11083		&676\\
$-18 \leq M_r \leq -16$	&72		&0.67	&11038		&746\\[3pt]
\hline
\multicolumn{5}{c}{In substructure}\\[2pt]
$-23 < M_r \leq -16$		&90		&0.23	&10454		&802\\
$-23 < M_r \leq -20$		&24		&0.23	&10527		&865\\
$-18 \leq M_r \leq -16$	&18		&0.26	&10553		&612\\[3pt]
\hline
\multicolumn{5}{c}{$\tau_\text{inf} > 1$ Gyr}\\[2pt]
$-23 < M_r \leq -16$		&249		&0.54	&10856		&552\\
$-23 < M_r \leq -20$		&67		&0.56	&10885		&537\\
$-18 \leq M_r \leq -16$	&64		&0.51	&10837		&576\\[3pt]
\hline
\multicolumn{5}{c}{$\tau_\text{inf} < 1$ Gyr}\\[2pt]
$-23 < M_r \leq -16$		&111		&0.76	&10916		&1222\\
$-23 < M_r \leq -20$		&29		&0.69	&11081		&1287\\
$-18 \leq M_r \leq -16$	&26		&0.79	&11196		&1190\\[3pt]
\hline
\end{tabular}
\caption{Dynamical properties of the populations identified in A\,2151. \label{A2151_pop_prop}}
\end{center}
\end{table}

Comparing Fig. \ref{A2151_acum_distr_func} and the properties in Table~\ref{A2151_pop_prop}, we can observe that red and early-infall members share the radial distribution, but the early-infall subset has recessional velocities more concentrated near the cluster value. The recent-infall galaxies have both larger radii and higher velocity dispersion in velocities with respect to all the other populations. Even if red and blue galaxies share similar values for the properties listed in Table~\ref{A2151_pop_prop} as a whole population, blue dwarfs have higher velocity dispersion than the red counterparts (third and sixth rows in Table~\ref{A2151_pop_prop}). 

The substructures are mainly located near the centre of the cluster. Moreover, they have larger velocity dispersion when are considered together as a unique subset. A detailed analysis of the properties of the three substructures is shown in Tab. \ref{A2151_sub_prop}. The S1 is the largest substructure in number of galaxies and it shows a velocity dispersion similar to the cluster value. The remaining two substructures, S2 and S3, have smaller velocity dispersions more similar to small galaxy groups. This complex substructure map indicates the young dynamical state of A\,2151 and hints a merging phase.

\section{The spectroscopic luminosity function of A\,2151}\label{lf_sec}
A good observable to analyse the influences of the environment and the physical processes on the cluster members is the LF. It has been extensively used with photometric data set, but only a few studies can be found with spectroscopic data down to the dwarf regime \citep{mobasher2003,rines2008,agulli2014,agulli2016a}. 

\begin{figure*}
\centering
 \includegraphics[width=1\linewidth]{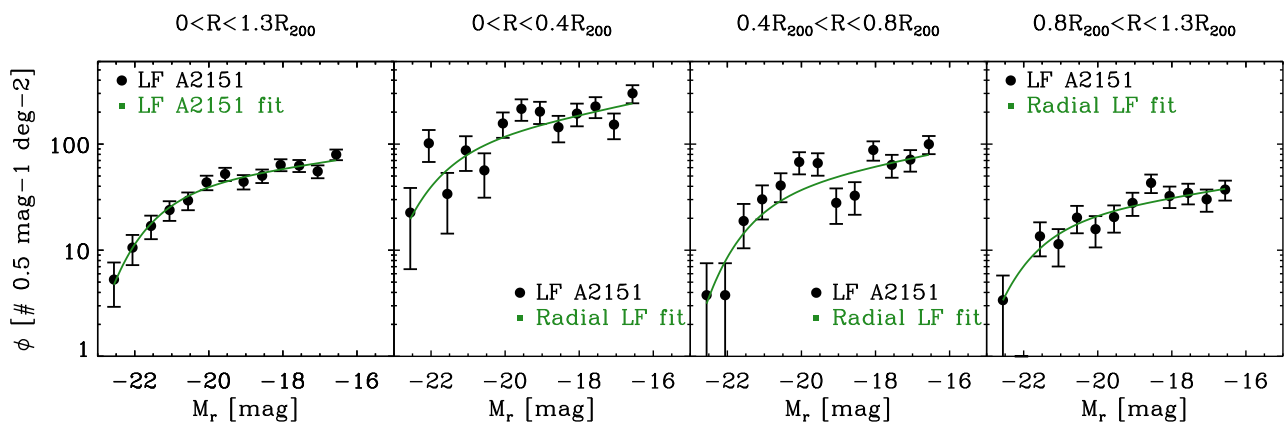}
 \caption{The black dots and their errors are the values of the LF of A\,2151 and the solid green line is the best fit, a single Schechter function. The first left-hand panel shows the global LF of the cluster. The other three panels present the LFs of the radial regions as reported on the top: the inner circle, the intermediate ring and the outer area.}
\label{A2151_lf}
\end{figure*}

Following the prescription in \cite{agulli2016a}, giving a certain binning of $M_r$ values, with $\Delta M_r = const = 0.5$\,mag, the LF can be computed as:
\begin{equation}\label{lf_eq}
\phi(M_r) = N_\text{phot}(M_r) \times  f_\text{m} (M_r) / (\Delta M_r \times A)
\end{equation} 
where $N_\text{phot}$ is the number of photometric targets (see section \ref{tg}), $f_\text{m}$ the fraction of cluster members defined in section \ref{memb}, and $A$ the surveyed area that correspond to a circle of $\sim 1.9$\,deg$^2$. We present the spectroscopic LF of A\,2151 reaching $M_r \sim -16$ in the left-hand panel of Fig. \ref{A2151_lf}, together with its best fit. The LF of this cluster is well modelled by a single Schechter function \citep{schechter1976}. This function has three free parameters: the normalisation, $\phi^*$, the characteristic luminosity of the cluster, $M^*$, and the slope, $\alpha$. In particular, $M^*$ represents the magnitude at which the Schechter function changes its dependence from an exponential to a power-law. The parameter $\alpha$ indicates the slope of the LF at faint magnitudes. Their values are given in Table~\ref{A2151_lf_param}.  


To study the effects of the low spectroscopic completeness at faint magnitudes on the LF of A\,2151, we used a stacked cluster obtained from the public database of the EAGLE cosmological hydrodynamical simulation \citep{mcalpine2016}. In this simulation, galaxies have been identified from the baryonic matter distribution and assigned a luminosity according to a stellar population synthesis model. The resulting LF is well modelled by a single Schechter function with $\alpha=-1.29$ and $M^*=-21.9$. We tested the solidity of our estimate of A\,2151 LF by drawing 10000 subsamples of this stacked cluster with the same spectroscopic completeness and magnitude range of our A\,2151 sample. The average parameters of the LF we obtain by analysing these subsamples with the same approach we used for  A2151 are $\alpha = -1.29 \pm 0.02$ and $M^* = -21.9 \pm 1.2$, therefore fully consistent with the expected values. We conclude that the spectroscopic completeness of the present study does not affect our results and discussion.

\subsection{Dependence of the LF on clustercentric distance}
This cluster does not have a BCG nor a concentrated X-ray emission in the centre. Therefore, the physical centre of the cluster is not so easy to determine. Moreover, the high fraction of galaxies in substructures hints a not virialized environment. However, the relation between galaxy density and clustercentric distance expected for nearby and almost virialized cluster is preserved. Therefore, we present a radial analysis of the LF. 

To study the relation of the LF with clustercentric distance, we divided the cluster into three regions, an inner circle of radius $R \leq 0.4 \, R_{200}$, an intermediate ring with $0.4 \, R_{200} < R \leq 0.8 \, R_{200}$ and the outer area with $0.8 \, R_{200} < R \leq 1.3 \, R_{200}$. We chose these intervals to have a similar and statistically significant number of members in each radial bin. The respective LFs and their best model functions are shown in Fig. \ref{A2151_lf}. The parameters of the single Schechter function used to fit the data are reported in Tab. \ref{A2151_lf_param}. 

A visual inspection of Fig. \ref{A2151_lf} suggests a dependency of the number density of the galaxies with the clustercentric distance. Indeed, the density is decreasing going from the centre to the outskirts of the cluster, as the normalization parameter traces with its decreasing values. However, the slopes of the LFs are similar. This result is highlighted by the $\alpha$ parameters which are compatible within the errors (see Tab.\ref{A2151_lf_param}). 

\begin{figure*}
\centering
 \includegraphics[width=1\linewidth]{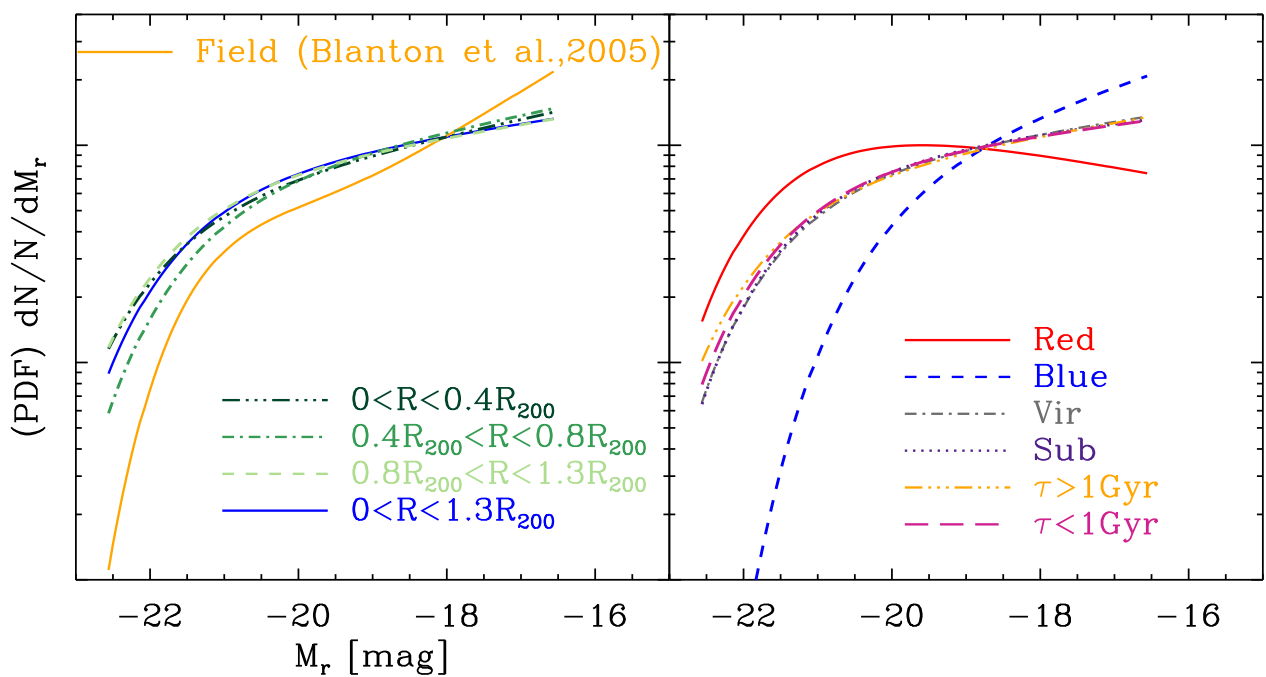}
 \caption{In the left-hand panel, we present the normalized radial LFs of A\,2151, from the inner to the outer regions with a green and line-style scale (see the legend), the global normalized LF with a blue solid line and the field spectroscopic and normalized LF by \protect \cite{blanton2005} with an orange solid line (The only one with an upturn). In the right panel we plot the normalized LFs of the different populations as described in the legend.}
\label{A2151_lf_norm}
\end{figure*}

We normalized the radial LFs using the integral of each LF -- i.e. within $\left[ -23 \leq M_r \leq -16 \right]$ -- as normalisation factor and we present the normalized curves in the left-hand panel of Fig. \ref{A2151_lf_norm}. We used this normalisation because it does not introduce any assumption of invariance in the study \citep[see the discussion in][]{agulli2016a}. A visual inspection of the figure hints compatible faint-end slopes of the LFs (see $\alpha$ in Tab. \ref{A2151_lf_param})  and a radial dependence of the  bright end that can be recovered in the different values of the $M^*$ parameter. However, this dependence is not linear. Indeed, the normalized LFs of the inner and outer regions of A\,2151 show an excess of bright galaxies and the brightest values of $M^*$ (see Tab. \ref{A2151_lf_param}), while the intermediate LF shows the faintest value. 

\begin{table}
\begin{center}
\begin{tabular}{ccc}
\hline
\hline
$\phi^*$		&$M_{r}^* $ [mag] 			&$\alpha$ \\
\hline
\multicolumn{3}{c}{$0 \, R_{200} < \,R\, \leq 1.3 \, R_{200}$}\\[5pt]
38.4		&$-21.80 \; _{-0.20}^{+0.22}$ 	&$-1.13 \; _{-0.02}^{+0.02} $\\[5pt]
\hline
\multicolumn{3}{c}{$0 < \,R\,\leq 0.4 \, R_{200}$}\\[5pt]
94.2		&$-22.13 \; _{-0.13}^{+0.13}$ 	&$-1.18 \; _{-0.01}^{+0.01} $ \\[5pt]
\multicolumn{3}{c}{$0.4 \, R_{200} < \,R\, \leq 0.8 \, R_{200}$}\\[5pt]
34.8		&$-21.67 \; _{-0.19}^{+0.21}$ 	&$-1.18 \; _{-0.01}^{+0.02} $\\[5pt]
\multicolumn{3}{c}{$0.8 \, R_{200} < \,R\, \leq 1.3 \, R_{200}$}\\[5pt]
18.9		&$-22.01 \; _{-0.32}^{+0.34}$ 	&$-1.14 \; _{-0.03}^{+0.03} $\\[5pt]
\hline
\end{tabular}
\caption{Schechter parameters of the total and the radial LFs of A\,2151. \label{A2151_lf_param}}
\end{center}
\end{table}

\subsection{The luminosity function of the galaxy populations}
To study the galaxy properties, we analysed the LFs of the different populations defined in Sec \ref{gc}. We will not perform a radial study of these subsets due to the poor statistics when divided in radial bins.

\begin{figure*}
\centering
 \includegraphics[width=1\linewidth]{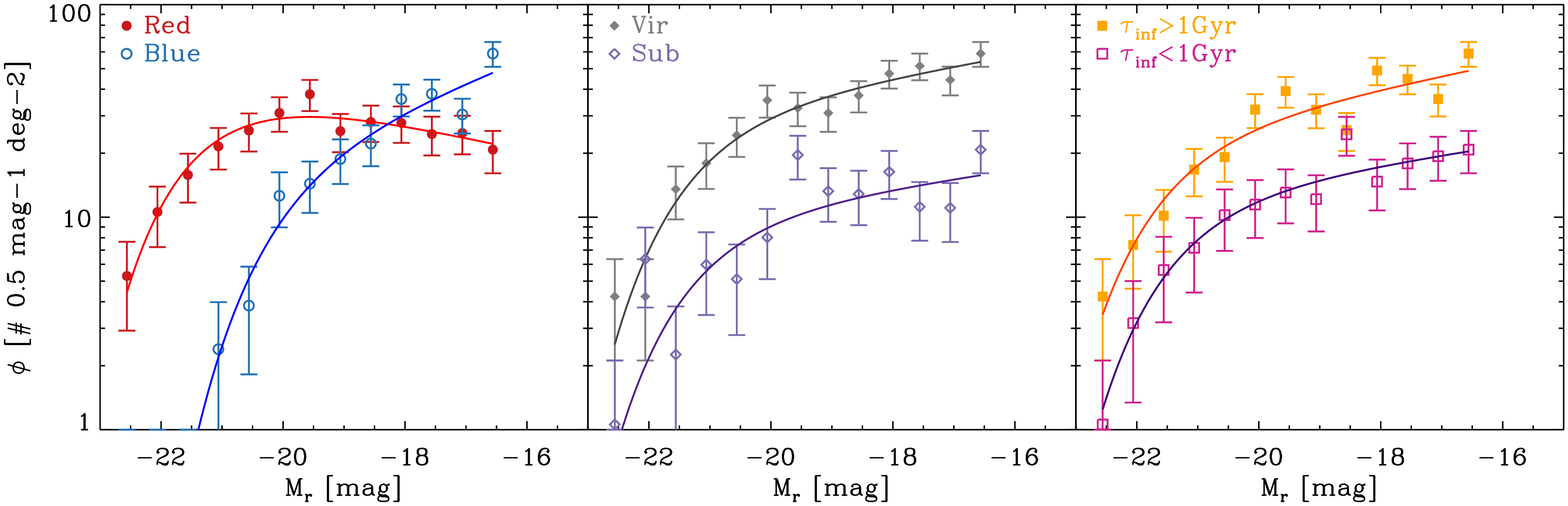}
 \caption{We present the LFs of the red (red dots) and blue (blue circles) population with their errors in the left-hand panel. Those of the virialized (filled grey diamonds) or in substructure (empty purple diamonds) and those of the early (filled orange square) and the late (empty magenta squares) infall members are shown in the middle and right-hand panels respectively. The solid lines are the best models: single Schechter functions.}
\label{A2151_lf_pop}
\end{figure*}

We show the resulting LFs in Fig. \ref{A2151_lf_pop}. All the populations can be modelled with a single Schechter and the values of the parameters are reported in Tab \ref{A2151_pop_param}. The comparison between the red and the blue populations is the most interesting. Indeed, while the virialized and early-infall members dominate the LF in the whole magnitude range, the red population leads only the bright part and the faint end, starting from $M_r \sim -18$, is dominated by blue galaxies. The $\alpha$ parameters highlight these results with values compatible within the errors for the virialized, in substructure, early- and late-infall LFs and a steeper $\alpha$ for the blue population. 

For an easier comparison of the overall shapes of these LFs, we normalized them using their integrals as normalisation factors and we show the normalized LFs in the right-hand panel of Fig. \ref{A2151_lf_norm}. A visual inspection hints that the stellar colour of the members is the physical property that strongly influences the probability density distribution of A\,2151.  Indeed, the other subsets based on dynamical properties, such as their infall time and/or virial equilibrium, have normalized LFs with invariant faint end slope and negligible differences at the bright end.  

\begin{table}
\begin{center}
\begin{tabular}{cccc}
\hline
\hline
Type 	&$\phi^*$		&$M_{r}^* $ [mag] 	&$\alpha$ \\
\hline
Red 		&46.6		&$-21.57 \; _{-0.22}^{+0.24}$ 	&$-0.84 \; _{-0.03}^{+0.04} $\\[5pt]
Blue 		&19.1		&$-20.32 \; _{-0.18}^{+0.19}$ 	&$-1.27 \; _{-0.04}^{+0.04} $\\[5pt]
\hline
Vir 		&30.1		&$-21.62 \; _{-0.22}^{+0.24}$ 	&$-1.13 \; _{-0.02}^{+0.03} $\\[5pt]
Sub 		&9.8			&$-21.59 \; _{-0.38}^{+0.40}$ 	&$-1.11 \; _{-0.05}^{+0.05} $\\[5pt]
\hline
$\tau_\text{inf} > 1$Gyr 	&24.1	&$-21.89 \; _{-0.26}^{+0.27}$ 	&$-1.15 \; _{-0.02}^{+0.03} $\\[5pt]
$\tau_\text{inf} < 1$ Gyr 	&12.3	&$-21.70 \; _{-0.35}^{+0.40}$ 	&$-1.11\; _{-0.04}^{+0.04} $\\[5pt]
\hline
\end{tabular}
\caption{Schechter parameters of the LFs of the populations identified in A\,2151. \label{A2151_pop_param}}
\end{center}
\end{table}

\subsection{LFs Comparison}\label{lf_cfr}
The power of the LF as an observable is the possibility of comparison with other studies. 
We will compare the LF of A\,2151 and the LFs of the red and blue populations with similar datasets of nearby clusters. 

We present the comparison among the normalized LFs of A\,2151 (this work), Abell 85 (A\,85) by \cite{agulli2016a}, Abell 2199 (A\,2199) and Virgo by \cite{rines2008} and the field by \cite{blanton2005}, where the normalisation factor is the integral of each LF. We selected these clusters because of the luminosity and the spatial coverage of these studies. Indeed, the similar magnitude limits and filters and the homogeneous mapping of the virial region make the comparison reliable. We also selected the photometric stacked LF by \cite{popesso2006}, normalized in the same way, because it is the most statistically significant example of the upturn that should be found in cluster LFs according to photometric predictions. The LF of A\,85 is the only one with spectroscopic data that seems to confirm the upturn, even if with a less steep slope \citep{agulli2014,agulli2016a}. 

\begin{figure*}
\centering
 \includegraphics[width=1\linewidth]{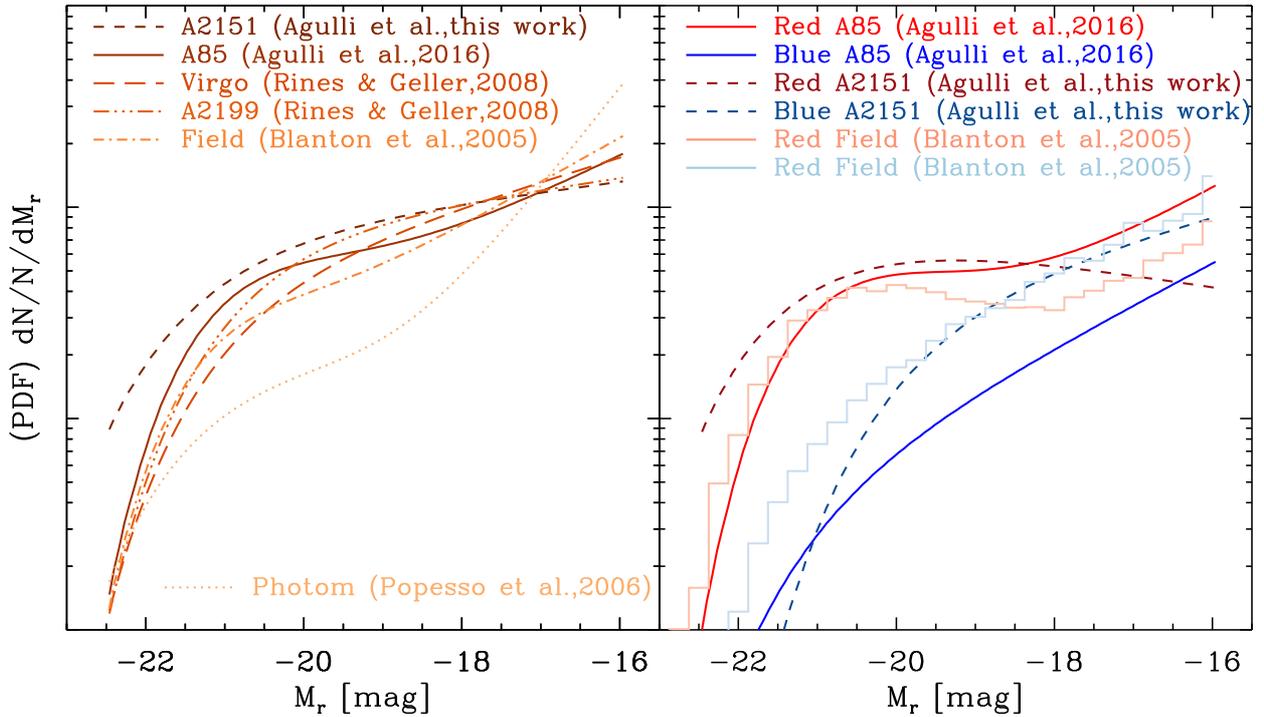}
 \caption{In the left-hand panel we present the normalised LFs of different nearby clusters (A\,2151, A\,85, A\,2199 and Virgo) with deep spectroscopic information, a normalized spectroscopic LF of the field and a normalized photometric stacked LF. For the reference studies, line-style and the colour gradient of the different LFs, we refer the reader to the plot legend. In the right-hand panel, we show the normalised LFs of the red and blue populations of A\,2151 with red and blue dashed lines respectively, the red and blue LFs of A\,85 with solid red and blue lines and the ones of the field populations with red and blue histogram styles.}
\label{LFs_cfr}
\end{figure*}

Fig \ref{LFs_cfr} presents the normalized LFs in the left hand panel. Even if A\,2151, A\,2199 and Virgo are all modelled by a single Schechter function, A\,2151 shows a higher probability of finding intermediate and bright galaxies when the normalized LFs are compared. 

Another interesting hint on galaxy evolution in different environment comes from the comparison of the LFs of the red and blue populations. We selected the field spectroscopic results by \cite{blanton2005} and the deep spectroscopic LFs of A\,85 by \cite{agulli2016a}. We normalized these LFs using the integral of the total LF of each cluster in order to enhance the relative properties of the two populations. A visual inspection of the normalized LFs presented in Fig \ref{LFs_cfr} suggests that the slopes at the faint end are compatible among the LFs, with the exception of the red subset of A\,2151 which shows a decreasing slope. A direct comparison of the values is not easy due to the different fits: the red LFs are modelled with a double Schechter function for A\,85 and the field and the blue LFs with a single Schechter function. However, we can observe that the bright part is always dominated by red galaxies; the LF of blue galaxies dominates the faint end both in the field and in A\,2151: in the field this occurs at $M_r \sim -19$, in A\,2151 one magnitude fainter, at $M_r \sim -18$.

\section{Discussion}\label{disc}
The literature on the LFs, and in particular the spectroscopic analysis, has pointed out an invariance of the bright end with respect to the environment \citep[e.g.,][]{andreon1998,goto2005,popesso2006,barkhouse2007,agulli2016a}. Indeed, in the nearby Universe, the colour of bright galaxies, and consequently their gas content, seems to be related to internal properties \citep[e.g.][]{tanaka2005,delucia2007,agulli2016b}. 

The results presented in this work are in agreement with previous conclusions. Indeed, the normalized LF of A\,2151 gives higher probability of having bright galaxies compared to other nearby clusters, and these bright galaxies are mainly red ones. This cluster is in its merging epoch and it has not formed a BCG yet. Therefore, we expect to find bright galaxies with a higher probability than in other clusters. Indeed, the cluster potential is not deep enough or the galaxies have not been there long enough to suffer of high dynamical friction. Moreover, the predominance of red bright galaxies is new evidence for the independence from the environment of the colour of luminous galaxies. The bright end of the LF is dominated by red objects for any environment considered in this study: (i) an almost relaxed cluster, A\,85, (ii) the field and (iii) a still forming cluster, A\,2151. 

Since we normalised to the total area, the higher probability of finding bright galaxies means a lower probability of finding faint ones. Indeed, when we compare the colour-magnitude diagram of this cluster with other clusters \citep[see e.g.][]{agulli2016b}, a clear lack of red dwarfs is visible. Moreover, the LF of the red population highlight the deficit of red dwarfs with its slope parameter ($\alpha = -0.84$). 

Recently, \cite{eardley2015} analysed the LF with the \textit{GAMA} data set and presented a significant variation in the LF between different environments, but not with respect to the cosmic web \citep{eardley2015}. They presented a statistical analysis of the local density of the three fields observed by the survey that identifys voids, sheets, filaments and knots. They obtained four LFs well represented by a single Schechter function. The fits show compatible slopes among the four LFs, and a variation of the $M^*$ parameter, being brighter for denser regions. However, their $M^*$ parameters are fainter than the value for A\,2151 and similar to those of the other clusters analysed in Sec. \ref{lf_cfr}.

\begin{figure*}
\centering
 \includegraphics[width=1\linewidth]{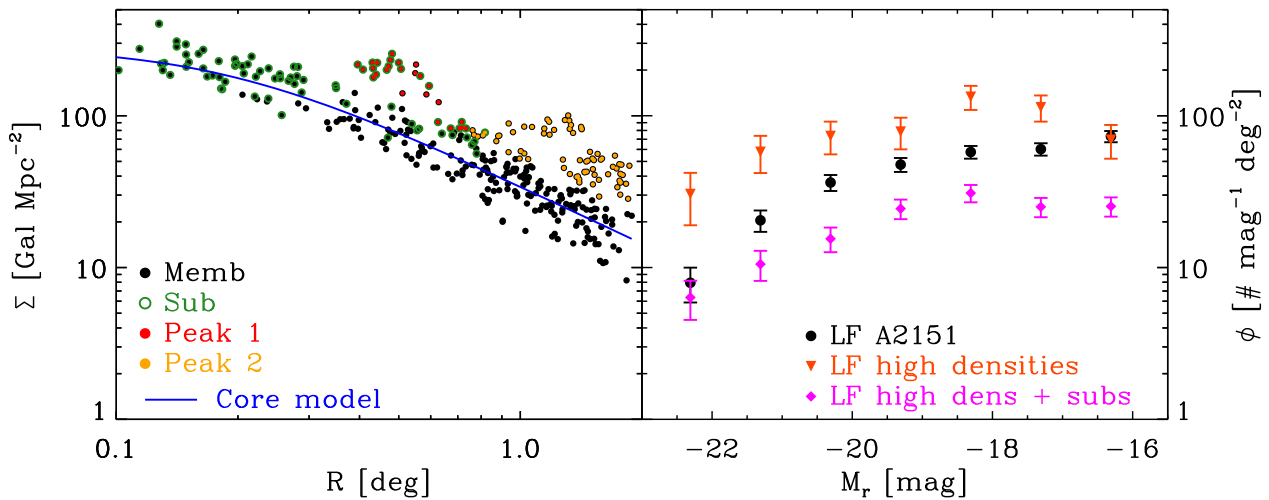}
 \caption{In the left-hand panel we present the projected local density of A\,2151 members (black dots) identifying the galaxies in substructures (green circles) and the two peaks of overdensities (red and orange dots). In the right-hand panel we present the LF of A\,2151 (black dots), the stacked LF of the two overdensities (dark orange triangles) and the stacked LF of the two overdensities and the substructures (magenta diamonds).}
\label{a2151_lf_hd}
\end{figure*} 

The presence of a dip in the LF and the change in the slope described by a double Schechter function opened a debate in the literature. The dip is thought to depend on a selective transformation of intermediate luminosity galaxies into brighter ones by dynamical friction which should be stronger in low-density groups \citep[e.g.][]{ostriker1980,trentham2002,miles2004,miles2006}. Moreover, trying to understand this effect, \cite{miles2004} made a \textit{toy model} of galaxy mergers in groups. They found that the lower the velocity dispersion of the group, the higher the evidence of evolution, with the presence of a dip in the LF. 

When a group enters in a cluster and survives, it can be seen as an overdensity in the projected local galaxy density profile. Therefore, \cite{lee2016} identified these peaks and obtained a LF with a more pronounced dip with respect to the LF of their cluster. Moreover, \cite{agulli2016a} estimated the LFs for samples at different distances from the cluster centre and observed a dip in the LFs of the intermediate and the external radial bins. A\,2151 is one of the few nearby clusters in its merging phase. 
However, the substructure analysis performed in the present work does not show any dip or change in slopes (se Fig.~\ref{A2151_lf_pop}). Therefore, we investigated the dependence of the LF on the local density. We present the projected local density profile and the stacked LF of the two overdensity peaks in Fig. \ref{a2151_lf_hd}. We used the luminosity bins of 1 magnitude instead of 0.5 because of the lower statistics. Considering only these two peaks or considering them together with the substructures identified with the caustic method (see Sec. \ref{sec_sub}) gave similar results: no dips and no changes in the slope. We observed that the velocity dispersions of the substructures (see Tabs \ref{A2151_pop_prop} and \ref{A2151_sub_prop}) or of these two peaks ($\sim 880$ and $\sim 630$\,km/s, respectively), are higher than the values usually considered for small groups. However, the absence of a dip in A\,2151 suggests that there are other processes that influence the evolution of faint galaxies in clusters. 

\cite{tully2002} suggested a relation between the shape of the LF and the mass of the cluster which was confirmed by \cite{zandivarez2011}. \cite{agulli2016a} suggested that the same relation explains the differences among the shapes of the deep spectroscopic LFs of the nearby clusters that they compared. With our data set, the $M_{200}$ of A\,2151 results to be $3 \times 10^{14}$\, M$_{\odot} \, h^{-1}$, higher than the masses of Virgo, A\,2199 and A\,85 \citep[$M_{200} = 2.0, 2.3, 2.4 \times 10^{14}$\, M$_{\odot} \, h^{-1}$, respectively,][]{rines2006}. 
Nevertheless, we need more deep spectroscopic observations of clusters spanning a wide range of masses to better constrain this relation.  
 
\begin{figure}
\centering
 \includegraphics[width=1\linewidth]{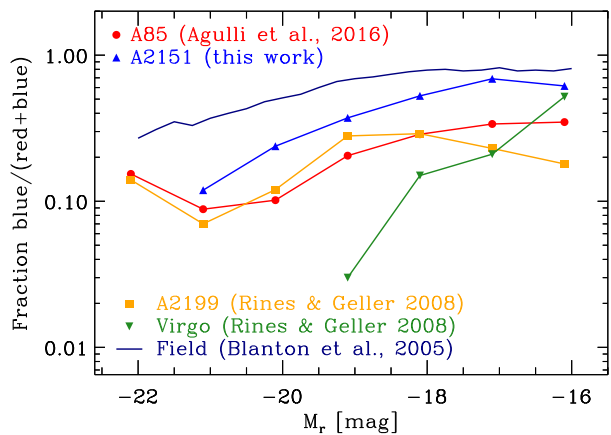}
 \caption{The blue upward triangles are the blue to total ratio for A\,2151 with the populations defined in this work. The red dots are the blue to total ratio for A\,85 obtained with the data by \protect \cite{agulli2016a}. The ratios for A\,2199 (orange squares) and Virgo (green downward triangles) are estimated using the data provided by \protect \cite{rines2008}.}
\label{rb_ratio_cfr}
\end{figure}

The evolutionary state of the cluster environment can also strongly affect galaxy evolution and in particular their star formation. Therefore, the differences observed in the LFs of these nearby clusters can depend on the red and blue populations \citep[][]{li2009}. Thanks to the public catalogues for A\,2199 and Virgo by \cite{rines2008}, we estimated their red and  blue populations following the prescription in Sec. \ref{red_blue}. We also have the two populations identified for A\,85 \citep{agulli2016a}. Therefore, we analysed the fraction of blue galaxies for the four clusters available in the literature with deep spectroscopy and we present them in Fig. \ref{rb_ratio_cfr}. In order to compare them, we applied the \textit{K}--correction given by \cite{chilingarian2012} to the absolute magnitudes. To compare our results with the field, we estimate the fraction of blue galaxies from the ratio of the blue and the total LFs by \cite{blanton2005}. 

At the bright end the clusters present similar behaviour except for Virgo, where no blue bright galaxies are observed. 
The visual inspection of Fig. \ref{rb_ratio_cfr} shows that the ratio of A\,2151 dominates at magnitudes fainter than  $M_r \sim -20$ and is similar to the field fraction at the low-mass end. Moreover, the fraction of Virgo increases with fainter luminosities. The global percentage of blue low-mass galaxies with respect to the  dwarfs is 25, 27, 27 and 66 per cent for A\,2199, A\,85, Virgo and A\,2151 respectively. Therefore, the evolutionary stage of the cluster clearly affect the star formation history.  Indeed, the one with the lowest fraction is a completely relaxed cluster, while the same fraction increases as the system goes out of the dynamical equilibrium. A transformation of blue dwarfs into passive red ones is expected in A\,2151 and will modify the faint end of the LFs of these two populations. However, to relate the star formation history estimated by the fraction of blue galaxies to the slope of the LF, a wider sample of clusters with deep spectroscopic information is needed.

 \begin{figure}
\centering
 \includegraphics[width=1\linewidth]{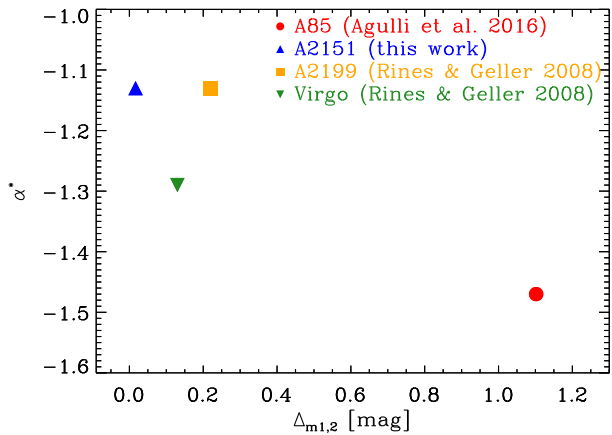}
 \caption{The faint-end slopes versus the magnitude gap between the brightest galaxies of A\,2151 (blue upward triangle), A\,85 (red dot), A\,2199 (orange square) and Virgo (green downward triangle). The errors are smaller than the symbols.}
\label{dm12}
\end{figure} 

The magnitude gap between the two brightest galaxies in a cluster could be an indicator of the dynamical state of the cluster. Therefore, we present the relation between this gap estimated as the simple difference between the two brightest members and the faint-end slope of the LFs discussed in this work in Fig. \ref{dm12}. A visual inspection suggests that the strongest difference is between A85, which is the only one presenting an upturn and requiring a double Schechter model, and the other three clusters. This relation is very interesting. However, the statistics is poor and a larger cluster sample is need to obtain a conclusive relation and to relate the differences observed in the LFs to the dynamical state of the cluster.

In the case of A\,2151, the peculiar dynamical state and merging history of the cluster can explain the flat faint end of the LF, and, in particular, the analogy between the LFs of the red and blue populations of the cluster and the field. Previous studies suggest that the environment may not play a key role in the number density of dwarfs, but only in their colour processing the gas of the faint galaxies \citep[e.g.,][]{tully2002,agulli2014}. Indeed, the shallow potential well of low-mass galaxies, together with physical mechanisms that process their gas such as tidal stripping, starvation, ram pressure or harassment, quenches the star formation and causes a reddening of the stellar populations \citep[e.g.,][]{quilis2000,bekki2002}. A\,2151 does not present the effects of these processes on its dwarf population suggesting that these galaxies have not been embedded in the cluster environment long enough to lose their gas or that the cluster environment itself is not yet hostile enough to act on their gas component. 

The radial analysis of the LF of A\,2151 shows that the faint end is independent of the clustercentric distance. The normalized LFs and the values of the $\alpha$ parameter highlight this result. However, the dependence on the radial distance from the centre is not completely understood in the literature. Indeed, cluster-to-cluster variations are observed. The spectroscopic mass function of an intermediate redshift cluster by \cite{annunziatella2016} confirmed the photometric findings of a clear steeping going inside-out, unlike nearby spectroscopic LFs \citep[e.g,][]{rines2008,agulli2016a}. However, even if the study of LF with deep spectroscopic data is powerful, only a handful of studies have been made so far. \cite{sanchez2005} presented a LF of A\,2151 with photometric data in the V magnitude. A direct comparison of the parameters is not possible due to the different magnitude band analysed. However, their trend of increasing faint-end slope for increasing clustercentric radius is not observed in our work. Beside the magnitude band, a possible explanation is the difference between photometric and spectroscopic analysis. Indeed, in the former, a statistical background subtraction must be applied and at the faint and denser regimes it can be underestimated. 

\begin{figure*}
\centering
 \includegraphics[width=1\linewidth]{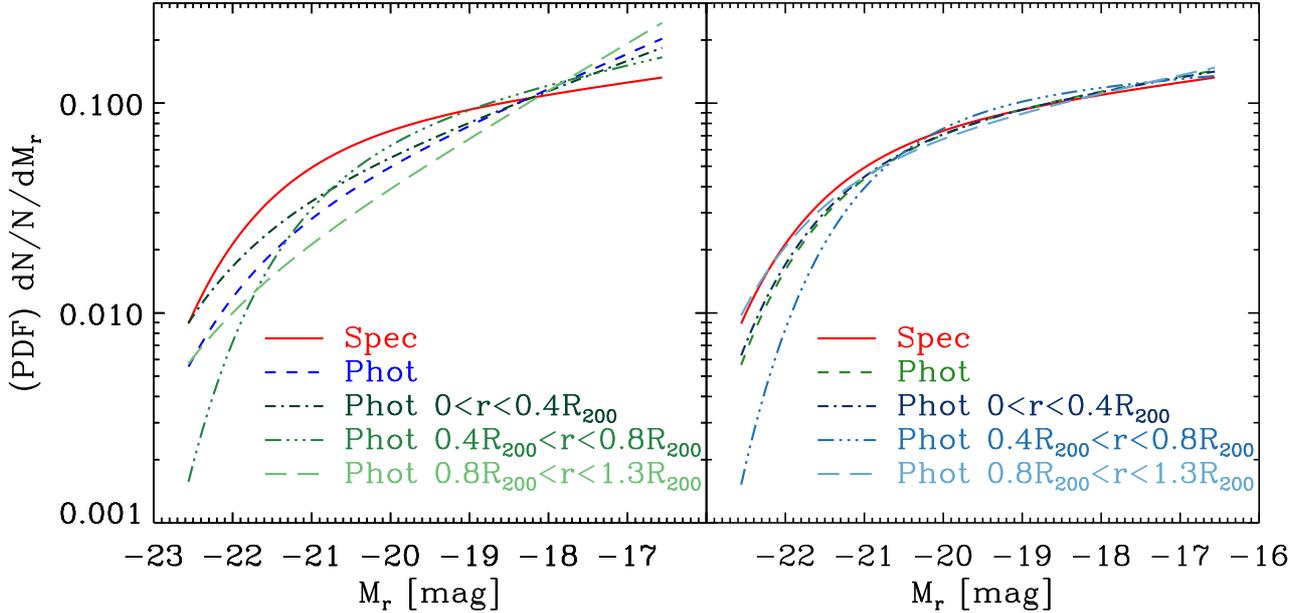}
 \caption{The normalized LFs of the spectroscopic (solid red line), the photometric (dashed blue - left-hand panel - or green - right-hand pnael), and the radial bins (green or blue colour scales in the left- and right-hand panels, respectively) are presented in this plot. In the left-hand plot there are the results obtained with the full catalogue. In the right-hand panel, the photometric LFs obtained after applying the colour cut are shown.}
\label{lf_phot}
\end{figure*} 

We estimated the global and the radial photometric LFs of A\,2151 as 
\begin{equation}
\phi_\text{phot}(M_r) = N_\text{phot}(M_r)/(\Delta M_r \times A) - \phi_\text{bkg} (M_r)
\end{equation}
where $N_\text{phot}$ is the number of photometric objects in our parent photometric catalogue and $\phi_\text{bkg} (M_r) = N_\text{bkg}(M_r)/(\Delta M_r \times A_\text{bkg})$ is the statistic estimation of the background. We used two different approaches. Firstly, we obtained the photometric LF from the total photometric catalogue and we used the background estimated by \cite{zarattini2015}. The resulting LFs, which have been normalised to their total integral, are presented in the left-hand panel of Fig. \ref{lf_phot}. With this analysis, a clear steepening of the faint end towards the outskirts is visible and the photometric LFs are steeper than the spectroscopic ones. In particular, the global LF has $\alpha = -1.41 \pm 0.02$.

With the second method, we applied the same colour cut we used for the target determination (see Sec.~\ref{tg}) to both the photometric catalogue and the background. In this case, we obtained the background  from 50 random regions taken from the SDSS database. We present the normalized LFs in the right-hand panel of Fig. \ref{lf_phot}. A visual inspection suggests no significant radial dependence or differences between the photometric ($\alpha = -1.15 \pm 0.02$) and the spectroscopic faint end slopes.

The comparison of the two panels of Fig. \ref{lf_phot} shows that applying the photometric LFs recover the radial trends and the steeper faint end observed in the literature. However, when we apply the colour cut to the catalogue and the background, no significant differences between the photometric and the spectroscopic LFs or among the radial bins are observed.


This cluster has a complex and clumpy matter distribution, which affects the identification of the cluster centre. Therefore, an other possible explanation for the distinct trends of the radial LFs between our work and the photometric one by \cite{sanchez2005} is the different centre selected.

\section{Conclusions}\label{concl}
The Hercules cluster presented in this work has 360 spectroscopically confirmed members within 1.3\,$R_{200}$, $M_r \geq -16$ and $\langle \mu_{e,r} \rangle < 23$ mag\,arces$^{-2}$. The dynamical analysis together with the different populations identified and the LFs show a peculiar nearby cluster. Indeed, the large fraction of galaxy in substructures, the absence of a BCG, the deficit of red dwarf galaxies observed in the LFs, the dwarf population dominated by blue galaxies together with deep X-ray analyses presented in the literature suggest that A\,2151 is a cluster in its merging epoch, still forming the dense and hostile cluster environment. Therefore, the internal or mass quenching and the dynamical friction drive the evolution of bright and intermediate galaxies. The evolution of the low-luminosity subset is dominated by environmental processes acting on their gas component, their stellar masses and their star formation histories. However, these processes are effective on long time scales and with hostile cluster environment.

\section*{Acknowledgements.}
We acknowledge the anonymous referee for the useful comments that helped us to improve the paper. We acknowledge Ana Laura Serra and Heng Yu for the helpful discussions on the results from the caustic method. We thank Claudio Dalla Vecchia for the helpful discussion on the EAGLE simulations and database. This work has been partially funded by the MINECO (grant AYA2013-43188-P). IA and AD acknowledge partial support from the INFN grant InDark. This research has made use of the Nineth Data Release of SDSS, and of the NASA/IPAC Extragalactic Database which is operated by the Jet Propulsion Laboratory, California Institute of Technology, under contract with the National Aeronautics and Space Administration. The WHT and its service programme are operated on the island of La Palma by the Isaac Newton Group in the Spanish Observatorio del Roque de los Muchachos of the Instituto de Astrof\'isica de Canarias.


\bibliographystyle{mnras}
\bibliography{/Users/ireagu/Documents/biblio/biblio} 

%
%


\appendix


\bsp	
\label{lastpage}
\end{document}